\documentstyle[psfig,aps,prl,twocolumn]{revtex}

\begin{document}

\draft \author{Markus Greiner, Immanuel Bloch, Olaf Mandel,
Theodor W.~H\"ansch$^*$, and Tilman Esslinger}
\address{Sektion Physik,
Ludwig-Maximilians-Universit\"at, Schellingstr.\ 4/III, D-80799
Munich, Germany and\\
Max-Planck-Institut f\"ur Quantenoptik, D-85748 Garching, Germany}
\title{Exploring phase coherence in a 2D lattice of Bose-Einstein condensates}

\maketitle
\begin{abstract}
\noindent Bose-Einstein condensates of rubidium atoms are stored
in a two-dimensional periodic dipole force potential, formed by a
pair of standing wave laser fields. The resulting potential
consists of a lattice of tightly confining tubes, each filled with
a 1D quantum gas. Tunnel-coupling between neighboring tubes is
controlled by the intensity of the laser fields. By observing the
interference pattern of atoms released from more than 3000
individual lattice tubes the phase coherence of the coupled
quantum gases is studied. The lifetime of the condensate in the
lattice and the dependence of the interference pattern on the
lattice configuration are investigated.
\end{abstract}


\noindent

Bose-Einstein condensates trapped in periodic optical potentials
provide a unique environment for experimental studies of a wide
range of physical phenomena
\cite{Kasevich98,Kasevich01,Jacksch98,Kutz01}. Here we use an
optical potential to create a two-dimensional periodic lattice
filled with several thousand 1D quantum gases. In the 1D quantum
gas the radial motion of the atoms is confined to zero point
oscillations and transverse excitations are completely frozen out.
In the degenerate limit, these systems are expected to show a
remarkable physics not encountered in 2D and 3D, for instance a
continuous crossover from Bosonic to Fermionic behaviour as the
density is lowered
\cite{Girardeau60,Lieb63,Olshanii,Petrov00,Girardeau01}. In our
two-dimensional periodic array of quantum gases the coupling
between neighboring lattice sites is controlled with a high degree
of precision by changing the intensity of the optical lattice
beams. After suddenly releasing the atoms from the trapping
potential we observe the multiple matter wave interference pattern
of several thousand expanding quantum gases. This allows us to
study the phase coherence between neighboring lattice sites, which
is remarkably long lived. Even for long storage times, when the
phase coherence between neighboring lattice sites is lost and no
interference pattern can be observed anymore, the radial motion of
the atoms remains confined to zero point oscillations.

\begin{figure}
\centerline{\psfig{file=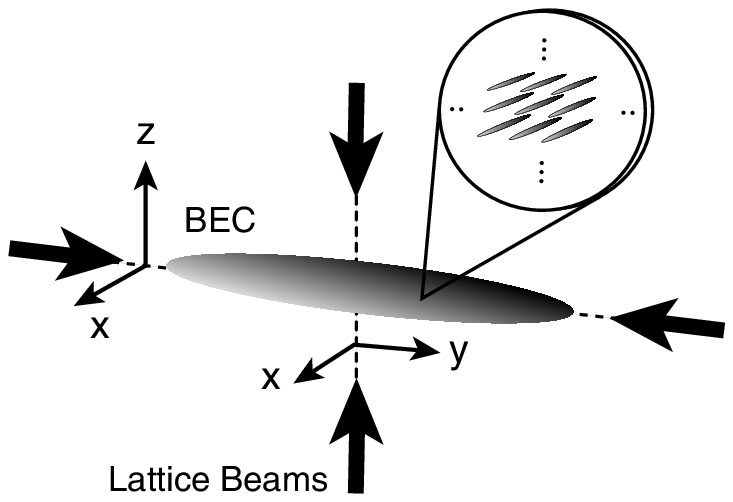,angle=0,width=0.3\textwidth}}
\caption{Schematic setup of the experiment. A 2D lattice potential
is formed by overlapping two optical standing waves along the
horizontal axis ($y$-axis) and the vertical axis ($z$-axis) with a
Bose-Einstein condensate in a magnetic trap. The condensate is
then confined to an array of several thousand narrow potential
tubes (see inset)}\label{fig:latticescheme}
\end{figure}

Similar to our previous work \cite{Greiner01}, almost pure
Bose-Einstein condensates with up to $5\times10^5$ $^{87}$Rb atoms
are created in the $|F=2,m_F=2\rangle$ state. The cigar shaped
condensates are confined in the harmonic trapping potential of a
QUIC-trap \cite{Esslinger98} with an axial trapping frequency of
24\,Hz and radial trapping frequencies of 220\,Hz. The lattice
potential is formed by overlapping two perpendicular optical
standing waves with the Bose-Einstein condensate as shown in
Fig.\,\ref{fig:latticescheme}. All lattice beams are derived from
the output of a laser diode operating at a wavelength of
$\lambda=852$\,nm and have spot sizes $w_0$ ($1/e^2$ radius for
the intensity) of approximately 75\,$\mu$m at the position of the
condensate. The resulting potential for the atoms is directly
proportional to the intensity of the interfering laser beams
\cite{Grimm00} and for the case of linearly polarized light fields
it can be expressed by:

\begin{eqnarray}
U(y,z) &=& U_0  \left\{\cos^2(k y) + \cos^2(k z) \right. \\
\nonumber && \left. + 2\,{\mathbf{e}}_1\!\cdot{\mathbf{e}}_2
\cos\phi \cos( k y) \cos(k z)\right\}. \label{eq:intensity}
\end{eqnarray}

Here $U_0$ describes the potential maximum of a single standing
wave, $k=2\pi/\lambda$ is the magnitude of the wave vector of the
lattice beams and ${\mathbf{e}}_{1,2}$ are the polarization
vectors of the horizontal and vertical standing wave laser fields,
respectively. The potential depth $U_0$ is conveniently measured
in units of the recoil energy $E_r = \hbar^2 k^2/2m$, with $m$
being the mass of a single atom. The time-phase difference between
the two standing waves is given by the variable $\phi$
\cite{Hemmerich}. In our setup this time-phase is measured
interferometrically and controlled with a piezo-mounted mirror and
a servo-loop. Furthermore, the intensity of the lattice beams is
stabilized in order to ensure a constant potential depth during
our measurements. The intensity pattern in the $y$-$z$-plane
extends along the $x$-direction, such that the resulting potential
can be viewed as a lattice of narrow tubes with a spacing of
$\lambda/2$ between neighboring lattice sites. These tubes provide
a tight harmonic confinement along the radial direction which
leads to large trapping frequencies $\omega_r \approx \hbar k
\sqrt{2U_0/m}$. In our setup potential depths of up to $12$\,$E_r$
are reached, resulting in a maximum radial trapping frequency of
$\omega_r \approx 2 \pi \times 18.5$\,kHz. The confinement along
the symmetry axis of a single tube is determined by the harmonic
confinement of the magnetic trap and the confinement due to the
gaussian intensity profile of the lattice laser beams. The
trapping frequencies along the symmetry axis of a single lattice
tube can be varied between $\omega_{ax} \approx 2\pi \times
10$\,Hz-$300$\,Hz. The spontaneous scattering rate due to the
lattice laser light is always less than $\Gamma_{sc} =
0.02\,\mbox{s}^{-1}$ and therefore negligible for our measurement
times.

In order to transfer the atoms into the lattice potential, the
laser power of the lattice beams is gradually increased in a
linear ramp to its final strength within 40\,ms. The atoms are
then held for a variable amount of time in the combined potential
of the interfering laser beams and the magnetic trap. The number
of occupied lattice sites can be estimated by counting the number
of lattice sites within the Thomas-Fermi extension of the
magnetically trapped condensate. For the above parameters we find
that up to 3000 lattice sites are populated, with an average
population of $\bar{N}_i \approx 170$ atoms per lattice site.

\begin{figure}
 \centerline{\psfig{file=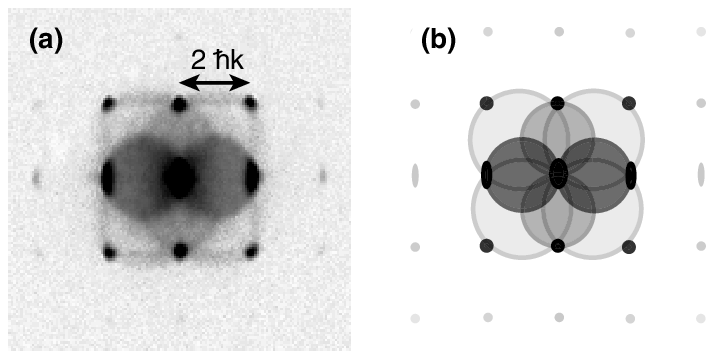,angle=0,width=0.43\textwidth}}
 \caption{(a) Average over 5 absorption images of released Bose-Einstein condensates
 that were stored in a 2D optical lattice potential. The maximum
 potential depth of the lattice was 12\,$E_r$ and the ballistic
 expansion time was set to 12\,ms. (b) Schematic image showing the
 expected discrete momentum states and the possible $s$-wave
 scattering spheres.}
 \label{fig:bildschema}
 \end{figure}

\begin{figure}
 \centerline{\psfig{file=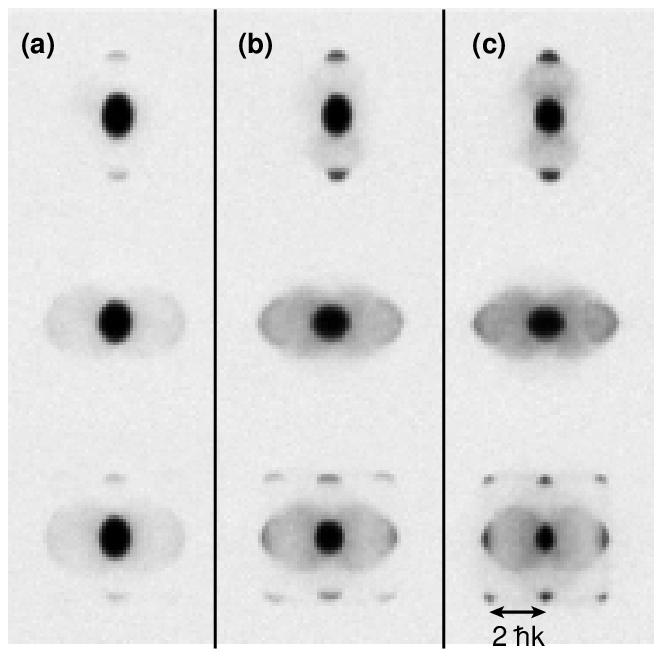,angle=0,width=0.28\textwidth}}
 \caption{Absorption images of Bose-Einstein condensates released from one-dimensional vertical (top row),
 one-dimensional horizontal (middle row) and two-dimensional horizontal+vertical (bottom row) lattice
 configurations. The images were taken for peak optical lattice depths of (a) 4\,$E_r$,
 (b) 8\,$E_r$, and (c) 12\,$E_r$.}\label{fig:kombibild}
 \end{figure}

When the atoms are released from the combined potential of the
optical lattice and the magnetic trap, the condensate wave
functions on different lattice sites expand and interfere with
each other. This interference pattern is imaged after a fixed
expansion time using absorption imaging, with the imaging axis
oriented along the $x$-axis and being parallel to the symmetry
axis of the individual lattice tubes. The results are displayed in
Fig.\,\ref{fig:bildschema} for a 2D-lattice potential with a
maximum potential depth of $12\,E_r$ and orthogonal polarization
vectors ${\mathbf{e}}_1\!\cdot{\mathbf{e}}_2=0$. In comparison,
Fig.\,\ref{fig:kombibild} shows the results for three different
potential depths of the optical lattice and 1D vertical
($z$-axis), 1D horizontal ($y$-axis) and 2D vertical+horizontal
lattice configurations (orthogonal polarization vectors
${\mathbf{e}}_1\!\cdot {\mathbf{e}}_2 = 0$). Several important
features can be seen on these images. First, discrete interference
maxima are visible that are arranged in a regular structure. These
interference maxima not only require a periodic density modulation
of the atoms but also phase coherence of the condensate wave
function throughout the lattice. They directly reveal the momentum
distribution of the atoms in the lattice. Second, $s$-wave
scattering spheres \cite{Band00,Ketterle00a} become more visible
as the higher order momentum components are more strongly
populated with increasing potential depth. These scattering
spheres occur due to collisions between atoms in the separating
momentum components after the trapping potential is switched off.
The collision probability between atoms in the horizontal momentum
components $|p_y|=2\,\hbar k$ and the $|p|=0$ momentum component
is high, due to the large extension of the condensate in the
horizontal direction. This yields long interaction times and thus
a high scattering probability. Along the vertical direction, the
size of the condensate is almost an order of magnitude smaller and
the interaction times are correspondingly shorter, resulting in a
much lower scattering probability. Furthermore $s$-wave scattering
spheres can also be seen in the diagonal direction due to
collisions between the diagonal momentum components and the
central $p=0\,\hbar k$ momentum component. For a maximum trapping
depth of $12\,E_r$ all of these 8 scattering spheres (see
Fig.\,\ref{fig:bildschema}(b)) are clearly visible in
Fig.\,\ref{fig:bildschema}(a).

The wave function of the Bose-Einstein condensate in the optical
lattice can be expressed as a sum of localized wave functions on
each lattice site. Such a localized wave function is described by
a gaussian wave function for the ground state of the tightly
confining radial axis of a single lattice tube, with radial widths
as low as 90\,nm. Along the weakly confining axis of a lattice
tube, the repulsive interactions between the atoms result in a
parabolic Thomas-Fermi profile with a maximum radial width of
$\approx5\,\mu$m. The maximum chemical potential per lattice tube
of $\mu \approx h  \times 6$\,kHz is then much smaller than the
radial energy level spacing, confining the radial atomic motion to
zero point oscillations.

\begin{figure}
 \centerline{\psfig{file=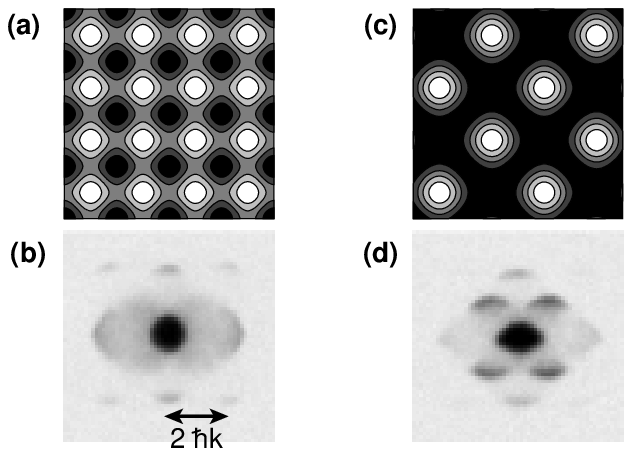,angle=0,width=0.35\textwidth}}
 \caption{Influence of the lattice configuration on the momentum distribution.
 For an optical lattice of (a) with orthogonal polarization vectors ${\mathbf{e}}_1\!\cdot{\mathbf e}_2 = 0$
 the first diagonal momentum orders with $|p|=\sqrt{2}\,\hbar k$ are
 suppressed (b) due to their vanishing geometrical structure factor.
 In contrast, if ${\mathbf{e}}_1\!\cdot{\mathbf e}_2 = 0$ and $\phi = 0$ as in (c), the resulting
 geometrical structure factor does not vanish for these momentum components
 and they are strongly visible (d).}
 \label{fig:structfact}
 \end{figure}

In addition to a strong dependence of the visibility of the higher
order momentum components on the localization of the wave
function, we find a suppression of momentum components due to
structural properties of the optical lattice. For a lattice
configuration with orthogonal polarization vectors between the two
standing waves for which the last term in the sum of
eq.\,\ref{eq:intensity} vanishes (see
Fig.\,\ref{fig:structfact}(a)), the first order diagonal momentum
components with $|p|=\sqrt{2}\,\hbar k$ are completely suppressed,
as can be seen in Fig.\,\ref{fig:structfact}(b). This is caused by
a destructive interference between matter waves emitted from
neighboring diagonal lattice planes and results in a vanishing
geometrical structure factor of these momentum components
\cite{Ashcroft}. If the lattice configuration is changed to
parallel polarization vectors between the two standing waves, such
that ${\mathbf{e}}_1\! \cdot {\mathbf{e}}_2 = 1$, and the
timephase is set to $\phi = 0$, the last term in
eq.\,\ref{eq:intensity} modifies the geometry of the lattice (see
Fig.\,\ref{fig:structfact}(c)). For this lattice configuration the
geometrical structure factor for the diagonal momentum components
with $|p|=\sqrt{2}\,\hbar k$ does not vanish and they are clearly
visible in the experiment (see Fig.\,\ref{fig:structfact}(d)).

\begin{figure}
 \centerline{\psfig{file=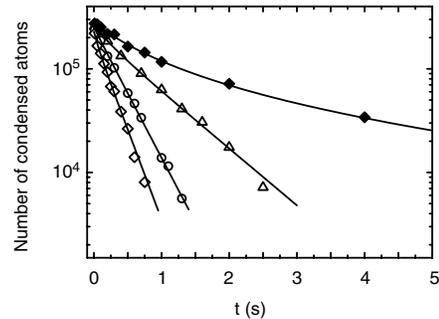,angle=0,width=0.32\textwidth}}
 \caption{Remaining number of condensed atoms after a variable hold time in the
 combined potential of the magnetic trap and the lattice potential (open datapoints) and in a pure
 magnetic trapping potential (solid diamonds). The maximum potential depth of the lattice was 4\,$E_r$ (open triangeles),
 8\,$E_r$ (open circles) and 12\,$E_r$ (open diamonds), respectively.}\label{fig:lifetime}
 \end{figure}

Measuring the number of condensed atoms after a variable hold time
in the optical lattice allows us to determine the lifetime of the
condensate in the lattice. For this, the lattice is ramped up to
its final strength within 40\,ms, then the atoms are held for a
variable amount of time in the lattice after which the lattice
potential is ramped down again in 40\,ms and the remaining number
of condensed atoms is measured after a ballistic expansion time
using absorption imaging. The slow ramp speed ensures that the
many-body wave function adjusts adiabatically to the changing
optical potential. The results of this measurement are displayed
in Fig.\,\ref{fig:lifetime} for a lattice configuration with
orthogonal polarization vectors ${\mathbf{e}}_1\!\cdot
{\mathbf{e}}_2 = 0$ (see Fig.\,\ref{fig:structfact}(a)). The
reduction of the lifetime of the condensate in the combined
potential of the magnetic trap and the optical lattice compared to
the lifetime in a pure magnetic trap is shown for three different
lattice potential depths. We believe this reduced lifetime to be
caused by remaining fluctuations of the lattice potential which
lead to a dephasing of the neighboring condensates with time. This
dephasing occurs faster with increasing potential depth, due to
the corresponding strong reduction in the tunneling matrix
element. In a band structure picture, the width of the energy
bands decreases strongly with increasing potential depth of the
lattice and external perturbations can then more easily lead to
nonadiabatic transitions within a single band, resulting in a
state where the atoms in a single lattice site remain within the
ground state of this lattice site, but do not exhibit phase
coherence to neighboring sites. Such dephased Bose-Einstein
condensates are not recombined into a single condensate when the
optical lattice potential is turned off adiabatically. This
strongly contributes to the measured decrease in condensate
fraction with increasing storage time.

\begin{figure}
 \centerline{\psfig{file=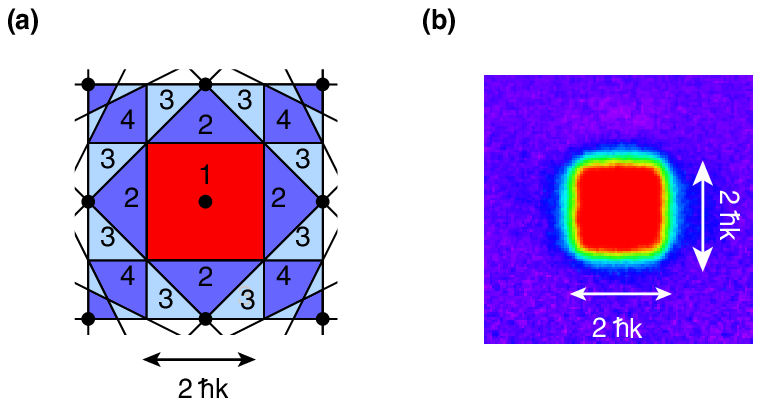,angle=0,width=0.42\textwidth}}
 \caption{(a) Reciprocal lattice and Brillouin zones for the two-dimensional Bravais lattice
 of Fig.\,\ref{fig:structfact}(a). (b)
 False color image of the experimentally measured band population of
 a dephased Bose-Einstein condensate in a 12\,$E_r$ deep optical
 lattice where phase coherence between neighboring lattice sites
 has been lost.}\label{fig:bandpop}
 \end{figure}

In order to verify this, we strongly reduce the tunneling matrix
element by turning off the magnetic trapping field and exposing
the atoms in a $12\,E_r$ deep optical lattice to the linear
gravitational potential. We then find that $2\,$ms after switching
off the magnetic field we do not observe an interference pattern
anymore after a sudden release of the atoms from the lattice
potential. This indicates that phase coherence of the atoms
throughout the lattice has been lost. In order to measure the band
population of such a dephased Bose-Einstein condensate, we ramp
down the optical potential in 2\,ms after a 2\,ms hold time in the
pure optical potential. This ramp speed ensures that we are
adiabatic with respect to the atomic motion in a single lattice
site and preserve the band population. We then let the cloud of
atoms expand for $12\,$ms and image the resulting momentum
distribution. Atoms originating from the first energy band are
then expected to obtain momenta that lie within the first
Brillouin zone of the lattice \cite{Kastberg95}. The Brillouin
zones of a two-dimensional Bravais lattice are shown in
Fig.\,\ref{fig:bandpop}(a), for which the first Brillouin zone is
a square of width $2\,\hbar k$. The experimental results in
Fig.\,\ref{fig:bandpop}(b) display a pronounced square-like
momentum distribution of width $2\,\hbar k$ which confirms that
the atoms from a dephased condensate populate only the first
energy band of the lattice. This directly proves that the atoms in
a single lattice tube remain in the radial ground state of the
system but no phase coherence between neighboring lattice sites is
observed anymore. The same method is used to measure the band
population after a variable storage time in the combined trapping
field of the magnetic trap and the optical lattice. We find that
for a $12\,E_r$ deep lattice 60\% of the initial number of atoms
are still present after a storage time of 1\,s and although no
significant condensate fraction can be measured for these storage
times (see Fig.\,\ref{fig:lifetime}) all of these atoms remain
confined to the first energy band, i.e. they are still in the
radial ground state of a lattice site. So far, we cannot
distinguish whether the atoms in a single lattice tube are also in
the axial ground state of this tube and only phase coherence
between neighboring lattice sites has been lost or whether
additional axial excitations in a single lattice tube are present.
Our measurements therefore yield a lower limit for the phase
coherence time of the atoms in the lattice and show that the
radial motion in a lattice tube remains confined to zero point
oscillations, i.e. is completely frozen out, even for long storage
times.

In conclusions we have studied the phase coherence properties of a
Bose-Einstein condensate stored in a two-dimensional optical
lattice potential. We have found that atoms from a Bose-Einstein
condensate can be efficiently transferred into the ground state of
the combined potential of the magnetic trap and the optical
lattice consisting of several thousand tightly confining potential
tubes, each filled with a 1D quantum gas. The phase coherence of
atoms throughout the lattice was found to be remarkably long
lived, while the radial motion of the atoms in a lattice tube
remained frozen throughout our measurement time.

By decreasing the density of the atoms per lattice site we are now
also able to enter the regime of quasi-condensates and even that
of a Tonks gas in which the fermionization of Bose systems
\cite{Girardeau60,Lieb63} can be studied. Such a Tonks gas
requires a very low number of atoms in a 1D geometry
\cite{Olshanii,Petrov00,Girardeau01} and it seems promising to
investigate this limit using the several thousand copies in our
2D optical lattice potential.\\

We would like to thank Wilhelm Zwerger and Martin Holthaus for
stimulating discussions and Anton Scheich for experimental
assistance during the construction of the experiment. We also
acknowledge support by the Deutsche Forschungsgemeinschaft.

\end{document}